\documentclass[aps,pre,twocolumn,groupedaddress]{revtex4-2}

\usepackage{amsmath}
\usepackage{graphicx}
\begin{document}


\title{Nonreciprocal Blume–Capel Model with Antisymmetric Single-Ion Anisotropies}


\author{Arjun R, Pratyush Prakash Patra, A. V. Anil Kumar}
\email[]{anil@niser.ac.in}
\affiliation{School of Physical Sciences, National Institute of Science Education and Research, Jatni, Bhubaneswar 752050, India}
\affiliation {Homi Bhabha National Institute, Anusaktinagar, Mumbai, India}


\begin{abstract}
We investigate the interplay between nonreciprocal interactions and chemical-potential imbalance in a two-species nonreciprocal Blume–Capel model. Combining a systematic mean-field bifurcation analysis with large-scale Monte Carlo simulations in two and three dimensions, we map the model’s dynamical regimes and transitions. Mean-field theory predicts a rich phase structure---disorder, a time-dependent “swap” (limit-cycle) phase, and static ordered states---separated by Hopf, saddle-node on invariant circle, saddle-node of limit cycles, pitchfork and saddle-node bifurcations. In two dimensions, Monte Carlo simulations reveal that spiral defects destabilise global swapping and, unless vacancies are strongly favoured, destroy long-range order. Crucially, a finite single-ion anisotropy $\Delta_\alpha = - \Delta_\beta$ promotes vacancy occupation in the $\alpha$ species and suppresses nonreciprocal dynamics, thereby restoring a robust static ordered phase. Finite-size scaling of susceptibility and Binder cumulants places the disorder $\to$ static transition firmly in the 2D Ising universality class. Moreover, within the static ordered phase, we observe a crossover that sharpens into a line of first-order phase transitions; these two regimes are separated by a critical point, analogous to the termination of the liquid–gas coexistence curve. In three dimensions, simulations largely mirror mean-field expectations, though swap $\to$ static ordering occurs indirectly via a disordered regime. Our results demonstrate that vacancy energetics provide a simple, experimentally relevant control knob that stabilises equilibrium-like order in nonreciprocal systems and that defects can generate novel critical behaviour.  
\end{abstract}


\maketitle
\section{Introduction}
Nonreciprocal interactions---i.e., interactions that violate action–reaction symmetry so that A’s influence on B differs from B’s influence on A---have recently emerged as a unifying framework for understanding the collective behaviour of a broad class of driven \cite{dzubiella2003depletion, golestanian2012collective, schmidt2019light}, active \cite{meredith2020predator, liebchen2017phoretic, cocconi2023active, dinelli2023non}, and open many-body quantum systems \cite{hatano1996localization, kawabata2022many, brighi2024nonreciprocal, wang2023quantum, li2024observation}.  Such interactions naturally arise in non-equilibrium settings, where gain, loss, or asymmetric couplings fundamentally violate microscopic reversibility. Concrete examples can be found across diverse disciplines, including neuroscience \cite{derrida1987exactly, sompolinsky1986temporal}, ecology \cite{romanczuk2009collective, barberis2016large, durve2018active}, population and social dynamics \cite{hong2011kuramoto, hong2011conformists, bunin2017ecological, caulkins2007explaining}. 
Beyond naturally occurring systems, nonreciprocity can also be deliberately engineered, for instance, in robotic swarms \cite{brandenbourger2019non, fruchart2021non, ghatak2020observation} and synthetic colloids \cite{lavergne2019group, bauerle2018self}. A defining consequence of nonreciprocity is the breakdown of detailed balance and, with it, the impossibility of describing the system in terms of a global free-energy function, which in turn permits a multitude of exciting phenomena and genuinely time-dependent macroscopic states that cannot be reduced to equilibrium ordering. Such time-dependent collective states include large-scale oscillations, travelling waves, and limit cycles of macroscopic order parameters \cite{avni2025dynamical, gupta2022active, kumar2025kinetic, you2020nonreciprocity, nasouri2020exact}, and raise the question of whether these behaviours should be regarded as bona fide non-equilibrium phases with sharp phase transitions and universal critical behaviour. Recent work on minimal spin models with macroscopic nonreciprocity has begun to address this question in a controlled setting. In particular, studies of the nonreciprocal Ising model \cite{avni2025nonreciprocal} have uncovered a rich phase diagram comprising a disordered phase, a static (ferromagnetic-like) ordered phase, and a dynamical “swap” phase characterised by persistent limit-cycle oscillations of the magnetisation. In three dimensions, the transition from disorder to the swap phase has been shown to be governed by a Hopf bifurcation, with static critical exponents consistent with the XY universality class, while in lower dimensions the stability and nature of the transition are strongly influenced by defect and droplet mechanisms. Remarkably, adding a diffusive random field to the system induces tricriticality, which manifests in mean-field theory as a Bautin (generalised Hopf) bifurcation \cite{kumar2025kinetic}. Many of the asymmetric systems that arise in nature can be mapped to nonreciprocal Ising \cite{avni2025dynamical} or similar models. Examples include models of neural network with asymmetric coupling \cite{sompolinsky1986temporal, sompolinsky2006theory, derrida1987exactly, amit1989modeling}, opinion dynamics and social systems \cite{fernandez2014voter, masuda2013voter, castellano2009statistical, bagnoli2015bifurcations, caulkins2007explaining}, driven non-identical parametric oscillators \cite{honjo2021100, goto2016bifurcation, alvarez2024biased, han2024coupled}. 

Spin models augmented by additional local degrees of freedom provide a powerful and analytically tractable framework for exploring collective phenomena in complex many-body systems. Among these, the Blume–Capel (BC) model \cite{blume1966theory, capel1966possibility} occupies a central role as a paradigmatic spin-1 lattice model in which each site can be either occupied by a spin taking values $S = \pm 1$ or remain vacant (S=0). The presence of a single-ion anisotropy term—often interpreted as a chemical potential for vacancies—introduces an additional control parameter that governs local occupation statistics and qualitatively enriches the phase diagram. Even in equilibrium, this extended local state space gives rise to phenomena absent in the spin-1/2 Ising model, including a tricritical point, first-order transitions, and vacancy-driven suppression of long-range order.

Generalising the Blume–Capel model to incorporate nonreciprocal, species-dependent onsite interactions offers a natural and minimally complex route to investigate how vacancy physics intertwines with nonreciprocal drive. In such a setting, imbalances in the effective chemical potential between different spin species compete with nonreciprocal couplings and intrinsic fluctuations, potentially reshaping both static ordering tendencies and dynamical instabilities. This extension, therefore, provides a controlled platform to explore how local vacancy energetics can either stabilise conventional ordered phases or promote genuinely non-equilibrium, time-dependent states. Beyond its intrinsic theoretical significance, the nonreciprocal Blume–Capel model (NR-BCM) constitutes a concrete minimal framework with direct applicability to a wide range of experimental, biological, and socio-physical systems. Representative realisations include networks of coupled nonlinear parametric resonators with cavities \cite{han2024coupled}, where asymmetric couplings and dissipation naturally give rise to effective nonreciprocal interactions.  In a biological context, coarse-grained models of interacting populations—such as predator–prey or competing-species systems with recruitment and extinction—map naturally onto the NR-BCM structure \cite{nareddy2020dynamical} (spin-1 variables and vacancy energetics) owing to population imbalance and the prey-predator nonreciprocal interactions.  These systems intrinsically exhibit nonreciprocal interactions due to asymmetric response or consumption dynamics between species. Similarly, spin-based models of neuronal networks, characterised by asymmetric synaptic couplings and unequal populations of excitatory and inhibitory neurons \cite{seung1997minimax, dominguez2000three, horiike2025distinct}, provide another natural setting in which NR-BCM–like physics can emerge. Extensions of opinion-dynamics models, such as the voter model \cite{fernandez2014voter}, which is itself fundamentally nonreciprocal, further illustrate the relevance of this framework when augmented to account for a nonparticipating or neutral subpopulation. Taken together, these examples underscore that the NR-BCM generalises the nonreciprocal Ising model, offering a versatile and unifying description that captures the interplay between nonreciprocity, local vacancy degrees of freedom, and collective non-equilibrium behaviour across diverse systems.

In this article, we investigate a two-species nonreciprocal Blume–Capel model with equal and opposing chemical potentials (rescaled) \(\Delta_A/k_BT=-\Delta_B/k_BT = \Delta \) and employ a combination of mean-field theory and large-scale Monte Carlo simulations to systematically characterise its steady-state behaviour. The mean-field analysis reveals a rich dynamical landscape comprising four distinct regimes: a disordered phase, a time-dependent “swap” phase, a statically ordered partial phase (only when $\tilde K = 0$), and a fully ordered phase. These regimes are separated by a hierarchy of bifurcations, including Hopf, saddle-node on invariant circle (SNIC), saddle-node of limit cycle (SNLC), pitchfork, and saddle-node bifurcations. In two dimensions (as observed in Monte Carlo simulations), the mean-field picture is profoundly altered by fluctuations and topological defects. In particular, spiral defects destabilise both the swap phase and the droplet-induced swap regime, while a finite single-ion anisotropy \(\Delta\) acts to stabilise the full ordered phase. The transition from disorder to full order is continuous and belongs to the two-dimensional Ising universality class. Moreover, within the full order phase, we observe a crossover which sharpens into a first-order transition at sufficiently large values of the effective coupling strength \(\tilde J\) (the first-order transition, however, does not lead to a new phase similar to the liquid-gas transition). This first-order line originates at a critical point. In three dimensions, the time-dependent swap phase remains stable \cite{avni2025dynamical}; however, when we vary $\Delta$, the transition from the swap phase to static order (which does not exist in the fully antisymmetric nonreciprocal Ising model \cite{avni2025nonreciprocal}) does not occur directly. Instead, we observe an intermediate disordered regime, leading to a sequence of transitions swap \(\to\) disorder  \(\to\) static order. Together, these findings demonstrate that species-dependent vacancy energetics provide a simple and effective control parameter that competes with nonreciprocal driving. By tuning this local chemical-potential imbalance, it is possible to suppress time-dependent collective dynamics and restore equilibrium-like static ordering in an otherwise intrinsically non-equilibrium, nonreciprocal system.

\section{Model}
The Blume-Capel model generalises the spin-1 Ising model by introducing a single-ion anisotropy (crystal-field) parameter $\Delta$, which controls the population of nonmagnetic states~\cite{blume1966theory,capel1966possibility}. The Hamiltonian is usually written as
\begin{equation}
{\cal H} = -J\sum_{<ij>}\sigma_i\sigma_j + \sum_{i} \Delta \sigma_i^2,
\end{equation}
where $J$ is the interaction strength. To introduce nonreciprocity into the model, we study a system containing two spin species, $A$ and $B$, with each lattice site $i$ allowed to host either species or both simultaneously. Spins of the same species interact reciprocally with their nearest neighbours, whereas spins of different species interact onsite nonreciprocally \cite{avni2025nonreciprocal}. Nonreciprocal systems are inherently non-Hamiltonian; hence, we define a “selfish” energy for the nonreciprocal Blume–Capel model. The selfish energy for the $i$-th spin of species $\alpha$ is defined as
\begin{equation}
    E_i^\alpha = -J\sum_{j_{nn}}\sigma_i^\alpha\sigma_{j_{nn}}^\alpha-K_{\alpha\beta}\sigma_i^\alpha\sigma_i^\beta + \Delta_\alpha(\sigma_i^\alpha)^2,
\end{equation}
where $\sigma_i^\alpha\in\{-1,0,1\}$, $J>0$ denotes the nearest-neighbour interaction strength between spins of the same species, and $\Delta^\alpha$ is the single-ion anisotropy (or chemical potential) for species $\alpha$; the sign of $\Delta^\alpha$ determines whether vacancies are favoured. The parameter $K_{\alpha\beta}$ encodes the nonreciprocal onsite interaction between different species. For a two-species system we set $K_{AB} = -K_{BA} = K > 0$, making $K_{\alpha\beta}$  fully antisymmetric. The system evolves through local spin updates in which each spin attempts to minimise its corresponding “selfish” energy \(E_i^\alpha\). We use the Glauber transition rate \cite{glauber1963time} for the spin update
\begin{equation} \label{\theequation}
    \omega_i^\alpha(\sigma_i^\alpha \to {\sigma_i^\alpha}') = \frac{e^{-\beta\Delta E_i^\alpha (\sigma_i^\alpha\to {\sigma_i^\alpha}')}}{\tau \sum_{p_i^\alpha}e^{-\beta\Delta E_i^\alpha (\sigma_i^\alpha\to p_i^\alpha)}},
\end{equation}
$\tau$ is the correlation time and $\Delta E_i^\alpha (\sigma_i^\alpha\to {\sigma_i^\alpha}')$ is the energy change. The dynamics of the system with $N$ lattice points is studied using the master equation \cite{van1992stochastic},
\begin{align}\label{\theequation}
    &\frac{\partial P(\sigma_1^A,\ldots,\sigma_N^A, \sigma_1^B,\ldots,\sigma_N^B; t)}{\partial t} = \nonumber\\ & -\sum_{i, \alpha}\left(\sum_{{\sigma_i^\alpha}' \neq \sigma_i^\alpha} \omega_i^\alpha(\sigma_i^\alpha\to{\sigma_i^\alpha}' )\right)P(\sigma_1^A,\ldots, \sigma_i^\alpha, \ldots,\sigma_N^B; t) \nonumber\\ & +  \sum_{i, \alpha} \omega_i^\alpha({\sigma_i^\alpha}'\to\sigma_i^\alpha)\left(\sum_{{\sigma_i^\alpha}' \neq \sigma_i^\alpha}P(\sigma_1^A,\ldots,{\sigma_i^\alpha}',  \ldots,\sigma_N^B; t)\right).
\end{align}
\noindent Here, $P(\sigma_1^A,...,\sigma_N^A, \sigma_1^B,....,\sigma_N^B; t)$ is the probability of finding the system at the configuration $\{\sigma_1^A,\ldots,\sigma_N^A, \sigma_1^B,\ldots,\sigma_N^B\}$ at time $t$.
\section{Mean-field analysis}
\begin{figure}[h!]
\includegraphics[width=1.0\linewidth]{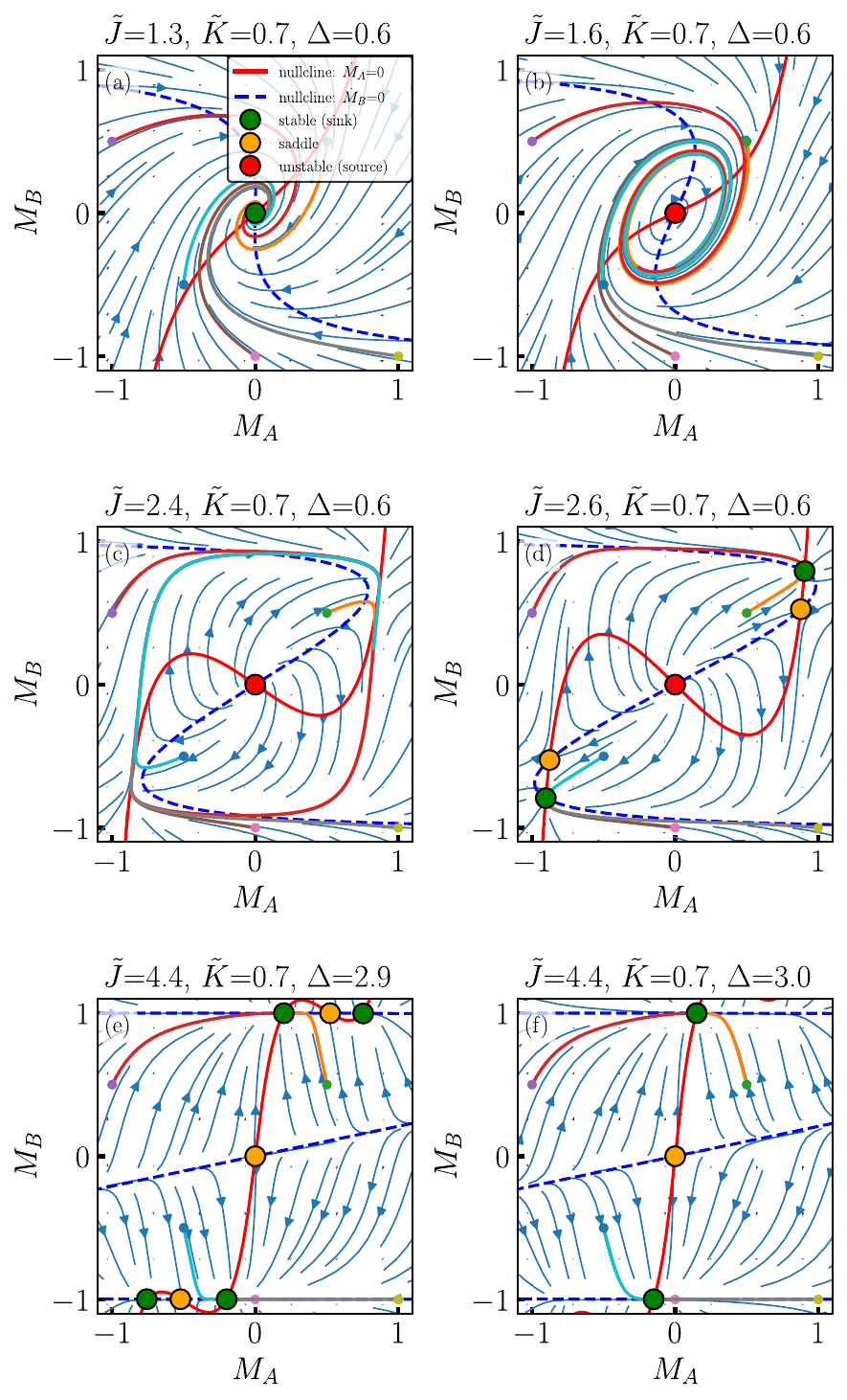}%
\caption{Phase portraits of the NR-BCM at nonzero nonreciprocity illustrating the bifurcations which take place as we vary the control parameters. (a) $\to$ (b): supercritical Hopf bifurcation at the origin. (c) $\to$ (d): saddle-node on invariant circle bifurcation. (e) $\to$ (f) saddle-node bifurcation of the saddle point and the stable fixed point with a higher magnitude of $R$. \label{3}}
\end{figure}

\begin{figure}[h!]
\includegraphics[width=1.0\linewidth]{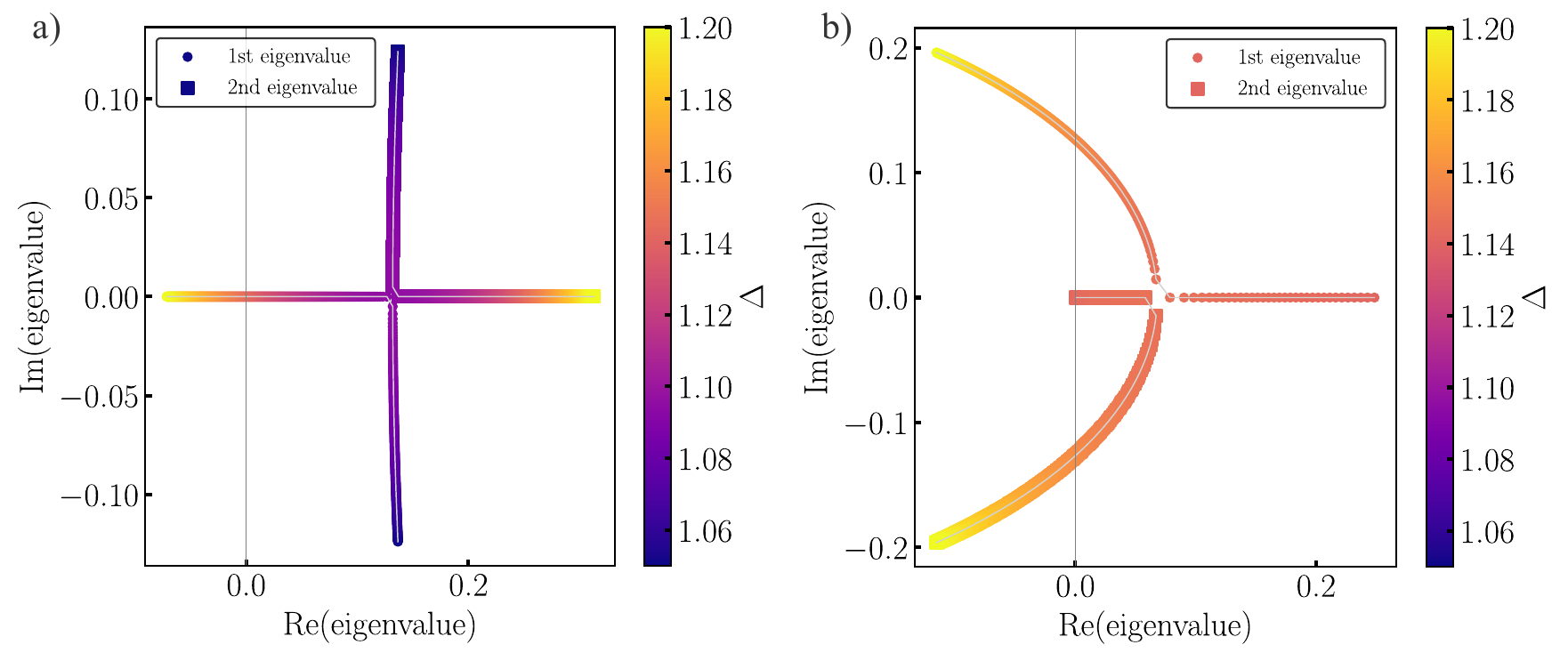}%
\caption{(a) Shows the eigenvalue flow of the trivial fixed point as we traverse from region II to III' at $\tilde J = 1.8, \tilde K= 0.7$. The flow indicates that for some value of $\Delta$, a subcritical pitchfork bifurcation occurs at the origin. (b) Shows the eigenvalue flow of the nonzero fixed points, which were created from the subcritical pitchfork bifurcation at the origin. As the magnitude of $\Delta$ increases, the eigenvalue crosses the imaginary line, corresponding to a subcritical Hopf bifurcation. \label{4}}
\end{figure}

The possible transition rates are
\begin{align}
    \omega_i^\alpha(-1\to-1) &= \omega_i^\alpha(0\to -1) = \omega_i^\alpha(1\to -1) = \omega_i^\alpha(-1)\nonumber\\ \quad & =\frac{\exp{(-x/k_BT)}}{\tau[2\cosh(x/k_BT) + \exp{(\Delta_\alpha/k_BT)}]}, \\
    \omega_i^\alpha(-1\to0) &= \omega_i^\alpha(0\to 0) = \omega_i^\alpha(1\to 0) = \omega_i^\alpha(0) \nonumber\\ \quad & =\frac{\exp{(\Delta_\alpha/k_BT)}}{\tau[2\cosh(x/k_BT) + \exp{(\Delta_\alpha/k_BT)}]}, \\
    \omega_i^\alpha(-1\to1) &= \omega_i^\alpha(0\to 1) = \omega_i^\alpha(1\to 1) = \omega_i^\alpha(1) \nonumber\\ \quad & =\frac{\exp{(x/k_BT)}}{\tau[2\cosh(x/k_BT) + \exp{(\Delta_\alpha/k_BT)}]},
\end{align}
where $x = J\sum_{j_{nn}}\sigma_{j_{nn}}^\alpha + K_{\alpha\beta}\sigma_i^\beta$. From \eqref{3} and \eqref{4}, we find the dynamics of the magnetisation
\begin{align}
    \tau\frac{d m_i^\alpha(t)}{d t} &= -m_i^\alpha(t) + \langle\omega_i^\alpha(1)\rangle - \langle\omega_i^\alpha(-1)\rangle.
\end{align}
Applying the mean-field approximation ($\langle g(x)\rangle = g(\langle x\rangle)$, 
assuming a spatially uniform magnetisation, and appropriately rescaling time and space \cite{avni2025dynamical}, we obtain
\begin{equation}\label{\theequation}
    \frac{d M_\alpha}{d t} = -M_\alpha + \frac{2\sinh{(\tilde{J}M_\alpha + \tilde{K}_{\alpha\beta}M_\beta})}{2\cosh{(\tilde{J}M_\alpha + \tilde{K}_{\alpha\beta}M_\beta)} + \exp(\Delta_\alpha/k_BT)},
\end{equation}
where $\tilde{J} = 2dJ/k_BT$ and $\tilde{K}_{\alpha\beta} = K_{\alpha\beta}/k_BT$. 

To differentiate between the phases of the system, we use the synchronisation ($R$) and the angular-momentum-like ($S$) order parameter.
\begin{align}\label{\theequation}
R &= \sqrt{\frac{M_A^2+M_B^2}{2}} \quad, & S &=  M_B\frac{\partial M_A}{\partial t} - M_A\frac{\partial M_B}{\partial t}.
\end{align}
The order parameters $R$ and $S$ quantify the synchronisation and the oscillations in the system, respectively \cite{avni2025dynamical}.  Also, the magnetisation vector $(M_A,M_B)$ will be used when we need to specify the exact macroscopic state. In our analysis, we focus specifically on the condition  $\Delta_A/k_BT = -\Delta_B/k_BT = \Delta$, which allows us to systematically explore the combined influence of population antisymmetry and fully antisymmetric nonreciprocal interactions.

According to Kolmogorov's criterion \cite{schnakenberg1976network}, a Markov process with rates 
$w(\vec{\sigma}\to\vec{\sigma}')$ satisfies detailed balance if, for any closed loop,
$$
\vec{\sigma}_1 \to \vec{\sigma}_2 \to \cdots \to 
\vec{\sigma}_n \to \vec{\sigma}_1,
$$
the forward and backward products of rates coincide:
\begin{equation}
\frac{\prod_{k=1}^{n} 
\omega(\vec{\sigma}_k \to \vec{\sigma}_{k+1})}
{\prod_{k=1}^{n} 
\omega(\vec{\sigma}_{k+1} \to \vec{\sigma}_k)} = 1 ,
\label{eq:kolmogorov}
\end{equation}
(with $\vec{\sigma}_{n+1}\equiv \vec{\sigma}_1$).  
Violation of Eq.~\eqref{eq:kolmogorov} for any loop implies broken detailed balance.

To demonstrate this in the NR-BCM,  
consider a single-site two-species state $(\sigma_i^A,\sigma_i^B)$ undergoing the
four-step cycle
$$
\uparrow\uparrow \to 
\uparrow\downarrow \to 
\downarrow\downarrow \to
\downarrow\uparrow \to 
\uparrow\uparrow,
$$
with all neighbours fixed in the up state.  
The single-spin-flip rates take the compact forms $f_1(x,y)=e^{-x}/(\tau(2\cosh x + e^y)),\qquad
f_2(x,y)=e^{x}/(\tau(2\cosh x + e^y))$. For this cycle, the product of forward and backward rates is
\begin{align}
W_{\mathrm{fwd}} &= 
f_1(\tilde J-\tilde K,-\Delta)\,
f_1(\tilde J-\tilde K,\Delta)\,
\nonumber\\[-2mm]
&\hspace{6mm}
\times f_2(\tilde J+\tilde K,-\Delta)\,
f_2(\tilde J+\tilde K,\Delta),
\\[1mm]
W_{\mathrm{bwd}} &= 
f_1(\tilde J+\tilde K,\Delta)\,
f_1(\tilde J+\tilde K,-\Delta)\,
\nonumber\\[-2mm]
&\hspace{6mm}
\times f_2(\tilde J-\tilde K,\Delta)\,
f_2(\tilde J-\tilde K,-\Delta).
\end{align}
Because all denominators cancel, the loop ratio reduces to the compact form
\begin{equation}
\ln\!\left(\frac{W_{\mathrm{fwd}}}{W_{\mathrm{bwd}}}\right)
= 8\tilde K .
\label{eq:cycle-affinity}
\end{equation}
Equation~\eqref{eq:cycle-affinity} is nonzero whenever $\tilde K\neq 0$, showing that the
Kolmogorov condition is violated, and detailed balance is broken.
The entropy produced per cycle is therefore
\begin{equation}
\frac{\Delta S}{k_B} = 8\tilde K ,
\end{equation}
demonstrating that the nonreciprocal coupling generates a finite cycle affinity
and drives the NR-BCM out of equilibrium.

\begin{figure}[h!]
\includegraphics[width=1.0\linewidth]{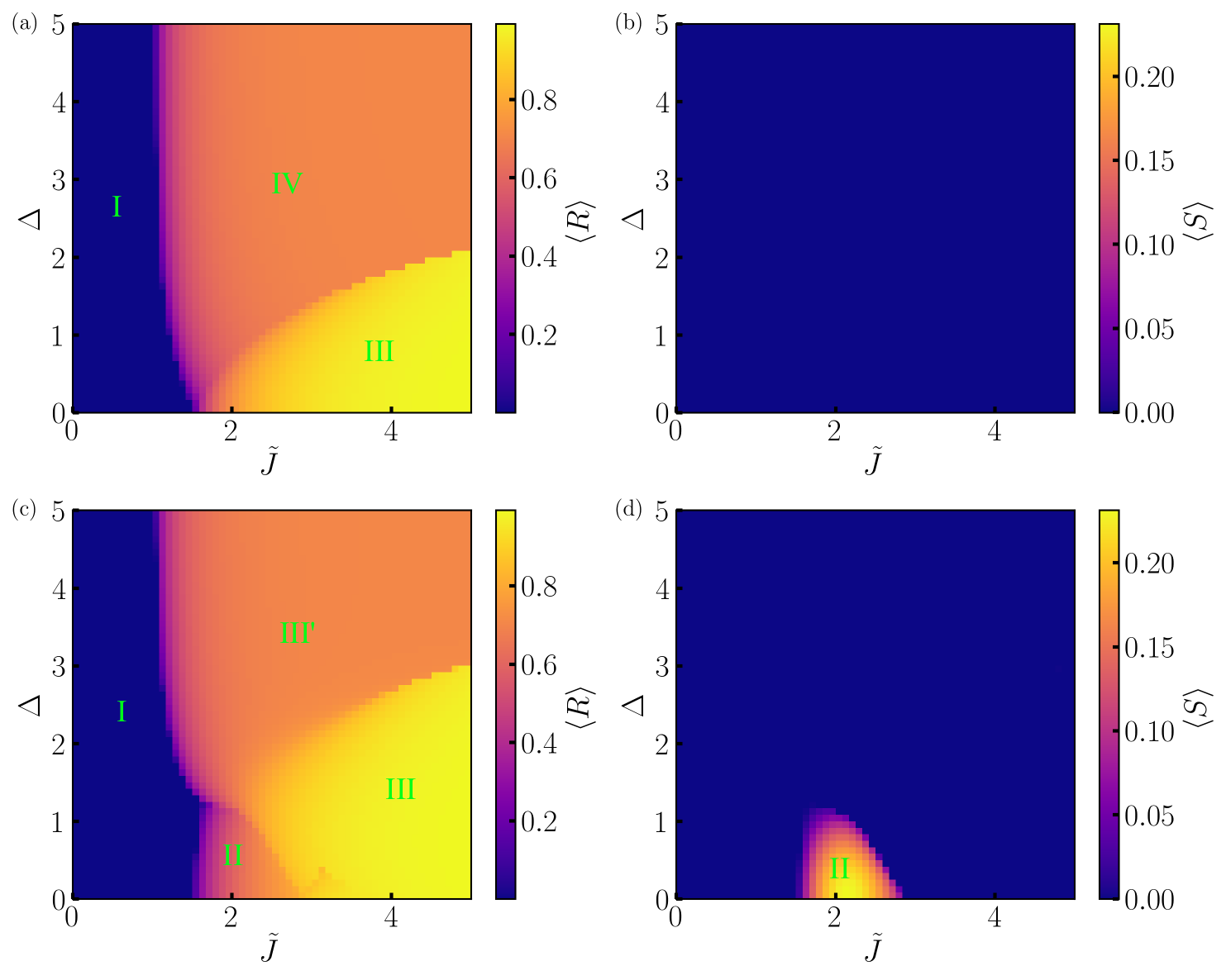}%
\caption{The figures show colour maps of $R$ and $S$. In the first row, $\tilde K = 0$ and it shows the colour map of (a) $R$ and (b) $S$. The second row corresponds to $\tilde K = 0.7$. (c) The colour map of $R$ for nonzero $\tilde K$ is topologically different than for $\tilde K = 0$. (d) Colour map of $S$ has nonzero values in the parameter space showing the presence of the swap phase. \label{1}}
\end{figure}

We then solve $\eqref{9}$ to construct the colour map shown in Fig.~\ref{1}, considering two distinct scenarios: one with zero nonreciprocity ($\tilde K$ = 0)  and the other with a finite non-zero nonreciprocity ($\tilde K \ne $ 0).  It is important to note that for any non-zero value of $\tilde K$, the overall topology of the phase diagram remains qualitatively the same, so the key effects can be captured by analysing a single representative non-zero $\tilde K$.  

Fig.~\ref{1}(a), (b) present the colour maps of the order parameters $R$ and $S$ in the absence of nonreciprocal interactions ($\tilde K$ = 0). Region~I denotes the disordered phase, characterised by vanishing magnetisations of both species ($M_A = 0, M_B = 0$). Region~III represents the fully ordered phase, in which both species develop nonzero magnetisation ($M_A \neq 0, M_B \neq 0$). Region~IV corresponds to a partially ordered phase, where only one species exhibits long-range order while the other remains disordered ($M_A = 0, M_B \neq 0$ or vice versa). The corresponding colour maps of $R$ and $S$ in the presence of finite nonreciprocity are shown in Fig.~\ref{1}(c), (d). In this case, Region~I again corresponds to the disordered phase, while Regions~III and III'  indicate fully ordered phases. A new dynamical regime, labelled Region~II, emerges due to nonreciprocal interactions. This region corresponds to the swap phase, in which the two species exhibit sustained time-dependent behaviour, periodically exchanging their magnetisations with a well-defined oscillation period.

To discuss discrete symmetries, we define the single-species inversions
\(\mathcal{I}_A:\sigma_i^A\mapsto-\sigma_i^A\), \(\mathcal{I}_B:\sigma_i^B\mapsto-\sigma_i^B\),
the global inversion \(\mathcal{I}_{AB}=\mathcal{I}_A\mathcal{I}_B\), and the species-exchange \(\mathcal{P}:(A\leftrightarrow B)\).
In NR-BCM, the selfish energy only consists of the global inversion symmetry \(\mathcal{I}_{AB}\) since it is even in spin variables, while single-species inversions $(\mathcal{I}_A$ and $\mathcal{I}_B$, as well as the label-exchange symmetry  $\mathcal{P}$ are no longer a symmetry due to the nonreciprocal coupling  \(K_{AB}=-K_{BA}\neq0\). These considerations determine the symmetries broken by the macroscopic states.

Within mean-field theory, the steady state self-consistency equations (of the fully antisymmetric NR-BCM) can be written schematically as
\[
M_\alpha = \mathcal{F}\big(h_\alpha,\Delta_\alpha\big),\qquad
\]
with effective fields \(h_A=\tilde J M_A +\tilde KM_B\) and \(h_B=\tilde J M_B-\tilde KM_A\).  The function ${F}(h,\Delta) = 2\sinh(h)/\big[2\cosh(h)+\exp{\Delta}\big]$ is odd in \(h\).  Imposing \(M_A=0\) while \(M_B\neq 0\) leads to the algebraic condition
\[
\mathcal{F}\big(\tilde KM_B,\Delta_A\big)=0.
\]
From the functional form of \(\mathcal{F}\), It is clear that the preceding equation can be satisfied only if \(\tilde KM_B=0\) (i.e. \(\tilde K=0\) or \(M_B=0\)), or that the system is taken to a singular limit in which the denominator of  \(\mathcal{F}\) diverges. A representative example of such a singular case is the limit \(\Delta_A\to\infty\) in the standard spin-1 parametrisation, where the species A becomes completely vacant. Aside from these scenarios, the condition could only be met through a non-generic, model-specific fine-tuned cancellation. Consequently, for any finite single-ion anisotropy $\Delta$ and nonzero nonreciprocity parameter ${\tilde K}$, a partial order state cannot exist in the mean-field dynamics. From a physical perspective, when one of the species is ordered, the inter-species coupling term $(\tilde K\sigma_i^\alpha)$ acts as a bias (which on average will have a nonzero magnitude when $\alpha$ is ordered) on the partner species and generically prevents a strictly partial-ordered solution at finite \(\Delta\).

The bifurcation structure of the mean-field flow is constrained by the aforementioned symmetries and is characterised by changes in the linear stability of fixed points \cite{kuznetsov1998elements}. A supercritical pitchfork bifurcation, in which a stable fixed point becomes unstable and gives rise to two new stable branches, generally corresponds to a continuous (second-order) phase transition in physical systems. Conversely, a saddle-node bifurcation involves the creation or annihilation of a pair of fixed points and acts as the mean-field signature of a discontinuous (first-order) transition. A line of saddle-node bifurcations can terminate at a cusp (or critical point) where a continuous transition emerges—analogous to a liquid–gas critical point—or it may intersect with lines of pitchfork or other saddle-node bifurcations. Finally, introducing nonreciprocity ($\tilde{K}_{AB} \neq \tilde{K}_{BA}$) breaks detailed balance; this generically facilitates Hopf bifurcations of symmetric fixed points, producing time-periodic (oscillatory or swap) phases.

\begin{figure}[h!]
\includegraphics[width=1.0\linewidth]{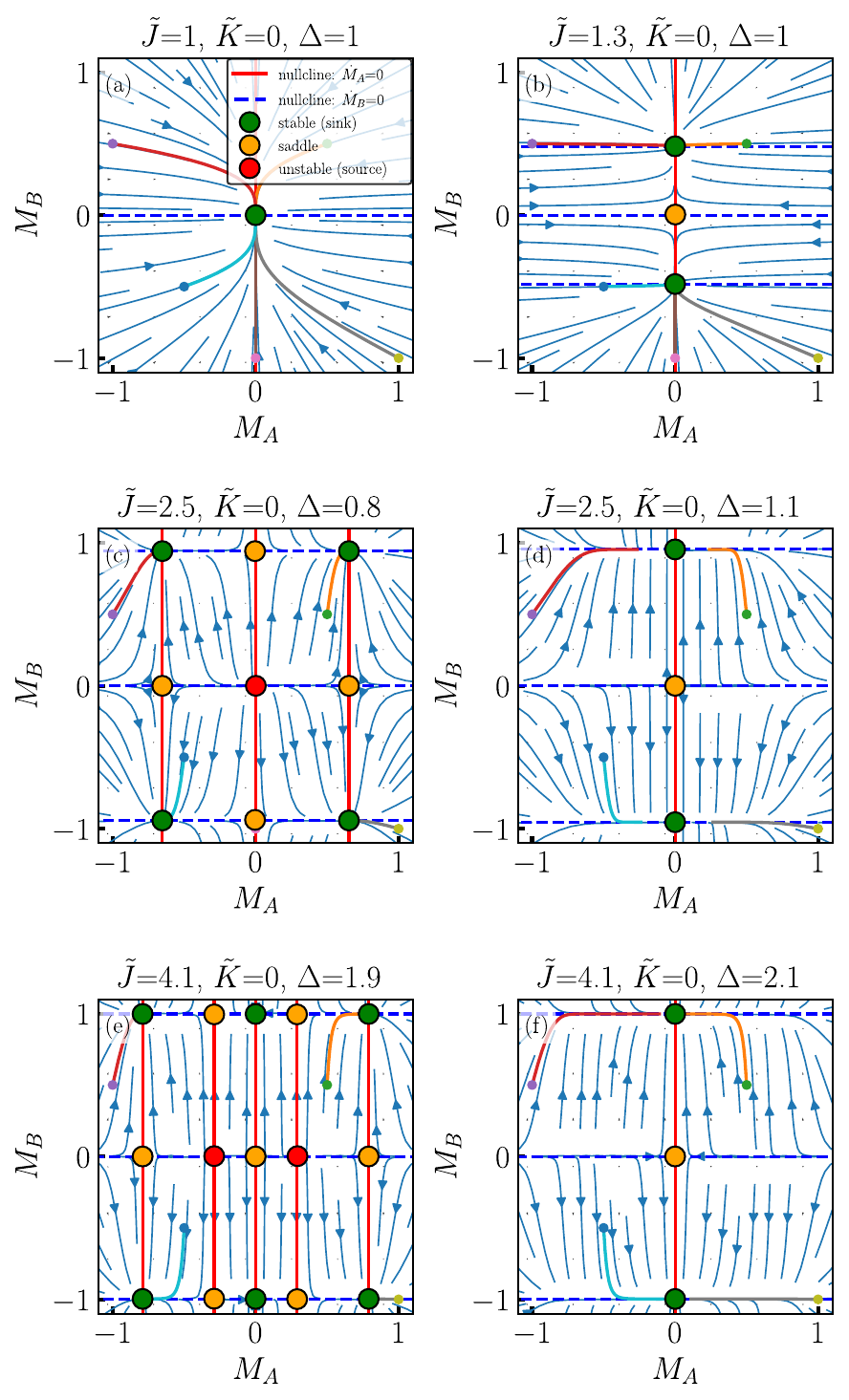}%
\caption{Phase portraits of the NR-BCM at zero nonreciprocity ($\tilde K = 0$) showing the bifurcations which take place as we vary the control parameters. (a) $\to$ (b): supercritical pitchfork bifurcation at the origin. (c) $\to$ (d): supercritical pitchfork bifurcation at the nonzero fixed points. (e) $\to$ (f) saddle-node bifurcation of the saddle point and the stable fixed point corresponding to the full order. \label{2}}
\end{figure}

In the absence of nonreciprocity, the system exhibits symmetry-breaking transitions driven by pitchfork and discontinuous transitions through saddle–node mechanisms. Specifically, the transition from region~I to IV (Fig.~\ref{1}(a)) proceeds via a supercritical pitchfork bifurcation at the origin $(M_A=0,\,M_B=0)$ (see Fig.~\ref{2}(a), (b)). For the transition between region~III and region~IV, the mechanism depends on the interaction strength $\tilde J$. For smaller values of $\tilde J$, this transition is mediated by supercritical pitchfork bifurcations at nonzero fixed points (Fig.~\ref{2}(c), (d)). However, for larger $\tilde J$, the transition changes character: a subcritical pitchfork bifurcation leads to the formation of one stable (partially ordered state) and two unstable nodes ($M_B > 0$ region in Fig.~\ref{2}(e); identical scenario takes place in the $M_B < 0$ region). The resulting unstable nodes subsequently undergo a saddle-node bifurcation, annihilating with the stable nodes associated with full order (Fig.~\ref{2}(f)). The lines representing the saddle-node and pitchfork bifurcations meet at a pitchfork degeneracy point, which corresponds to the tricritical point.

Introducing nonreciprocity qualitatively reshapes the bifurcation structure (Fig.~\ref{1}(c),(d)). Here, the partially ordered phase cannot exist, for instance, when species B is ordered, there is always an average non-zero bias on species A due to the inter-species coupling $\tilde{K}$. Therefore, the pitchfork bifurcation line disappears from the phase diagram. In contrast, the saddle–node line remains and terminates at a cusp bifurcation, analogous to a liquid–gas first-order line ending at a critical point. Moreover, for nonzero nonreciprocity, a distinct \emph{swap} phase emerges. The transition from the I (disordered state) into the II (swap phase) occurs through a continuous supercritical Hopf bifurcation (Fig.~\ref{3}(a), (b)), signalling the emergence of a stable limit cycle and the onset of sustained oscillatory behaviour. Subsequently, transitions from II to the fully ordered states in region III take place via a saddle-node on invariant circle (SNIC) bifurcation (Fig.~\ref{3}(c), (d)), highlighting the complex interplay between static ordering and time-dependent dynamics. The transition from III to III' is a crossover (no bifurcation) for the lower values of $\tilde J$ and a saddle-node bifurcation (Fig.~\ref{3}(e), (f)) otherwise. The transition from II to III' is nontrivial; there are multiple bifurcations in between which eventually end with a SNLC bifurcation.

To elucidate the stability and bifurcations within the transition from II to III', we examine the eigenvalue trajectories of the fixed points in the complex plane as a function of the parameter $\Delta$. At $\Delta = 1.05$ ($\tilde J = 1.8, \tilde K= 0.7$), a single fixed point exists at the origin, characterised as an unstable spiral locally  (Fig.~\ref{4}(a)) (corresponding to the swap phase). As $\Delta$ increases, the eigenvalues of this trivial fixed point converge onto the real axis, transforming the unstable spiral into an unstable degenerate node. A further increase in $\Delta$ causes the eigenvalues to split along the positive real axis, resulting in an unstable node. Eventually, one eigenvalue becomes negative, indicating a transition to a saddle point. This change is the result of a subcritical pitchfork bifurcation occurring at the origin, creating two new non-trivial fixed points. Initially, these new points possess positive, real eigenvalues, identifying them as unstable nodes (Fig.~\ref{4}(b)). As $\Delta$ continues to rise, these eigenvalues converge to form degenerate nodes before becoming complex with positive real parts, representing unstable spirals. Subsequently, the eigenvalues cross the imaginary axis, signifying a subcritical Hopf bifurcation at the non-trivial fixed points. Crucially, these events occur within the bounds of a larger limit cycle. The newly created unstable limit cycles expand with increasing $\Delta$ until they undergo a symmetric figure-eight homoclinic bifurcation. Further increase in $\Delta$ leads to the formation of an unstable limit cycle that surrounds all three fixed points \cite{kuznetsov1998elements}, which finally collides with the outer stable limit cycle, resulting in the mutual destruction of the oscillations via a saddle-node bifurcation of limit cycles. Consequently, the system remains with a saddle point at the origin and two stable spirals at the non-trivial fixed points. All of these bifurcations and phase boundaries are summarised in the phase diagram presented in Fig.~\ref{10}(a). For simplicity, in the nonzero nonreciprocity regime, we will refer to the fully ordered phase as static order.

\begin{figure}[h!]
\includegraphics[width=1.0\linewidth]{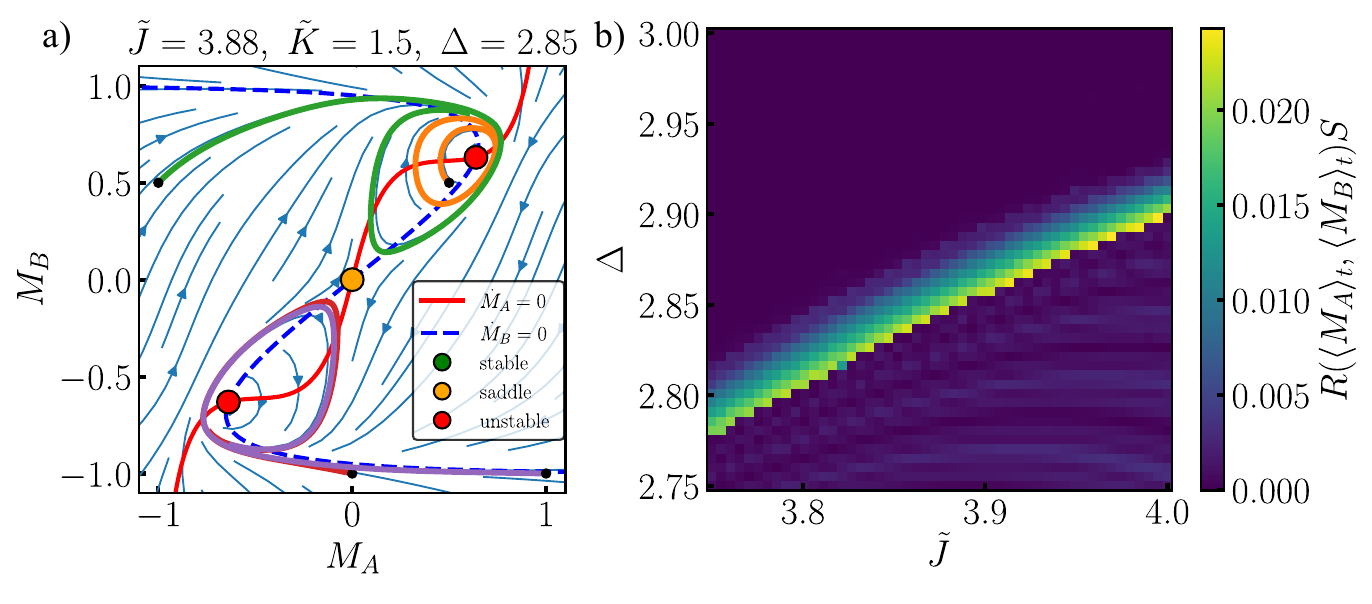}%
\caption{(a) Phase portrait illustrating the stable limit cycles at the nonzero fixed point. (b) Shows the colour map of $R(\langle M_A\rangle_t, \langle M_B\rangle_t)S$. Limit cycles centred at nonzero fixed points are evident from the map. Plotted at $\tilde K = 1.5$, $\Delta_B/k_BT = 0.7$ and scaling $\Delta_A/k_Bt = \Delta$ \label{5}}
\end{figure}

It is worth mentioning that in certain regions of the parameter space, we observe limit cycles centred at nonzero fixed points (see Fig.~\ref{5}(a)). This region has a null intersection with the region $\Delta_A/k_BT = -\Delta_B/k_BT = \Delta$. For instance, by fixing \(\Delta_B/k_BT=0.7\) and varying \(\Delta_A/k_BT=\Delta\), reveals this behaviour. However, the Hopf bifurcation, or the associated limit cycles around a nonzero fixed point of the system, cannot be independently captured by the order parameters $R$ and $S$. To better visualise these dynamics, we modify the order parameter $S$ by multiplying it with $R(\langle M_A\rangle_t, \langle M_B\rangle_t)$ where $\langle . \rangle_t$ denotes a time average. With this definition, $R(\langle M_A\rangle_t, \langle M_B\rangle_t)S$ vanishes for limit cycles centred at zero, allowing the detection of oscillations with nonzero centres. Using this approach, Fig.~\ref{5}(b) confirms the presence of limit cycles with nonzero centres (in the mean-field approximation), demonstrating the rich dynamical behaviour induced by nonreciprocal interactions.

\section{Monte Carlo Simulations}
We begin by performing two-dimensional Monte Carlo simulations of the NR-BCM on a square lattice with periodic boundary conditions. Spin updates are carried out using the Glauber transition rate given in equation \eqref{3}. Unless stated otherwise, each 2D simulation is run for $4\times10^6$ Monte Carlo sweeps, of which $ 2\times 10^5$ sweeps are used for equilibration (greater than the sweeps needed for $R$ to reach a steady state). Estimates of the critical exponents are obtained by averaging over 200 independent realisations, while all other observables are averaged over 20 realisations.

We also perform three-dimensional Monte Carlo simulations on a cubic lattice with periodic boundary conditions. The total number of Monte Carlo sweeps and the equilibration time are set to $4\times 10^6$ $\And$  $2\times 10^5$ respectively.  In this case, physical quantities are averaged over a single realisation, as our goal is primarily to illustrate the qualitative similarities between the 3D NR-BCM and the predictions of mean-field theory (while noting the expected deviations, which we discuss later in the paper).

For both plotting and peak detection, we used cubic-spline interpolation.  In practice, the susceptibility curves shown in the figures were smoothed with a cubic spline (natural boundary conditions) and interpolated on a dense grid; the interpolated maxima (positions and heights) were used to locate $\chi_{\max}$ for each system size and thereby to perform the finite-size scaling fits that yield the critical exponents.

\begin{figure}[h!]
\includegraphics[width=1.0\linewidth]{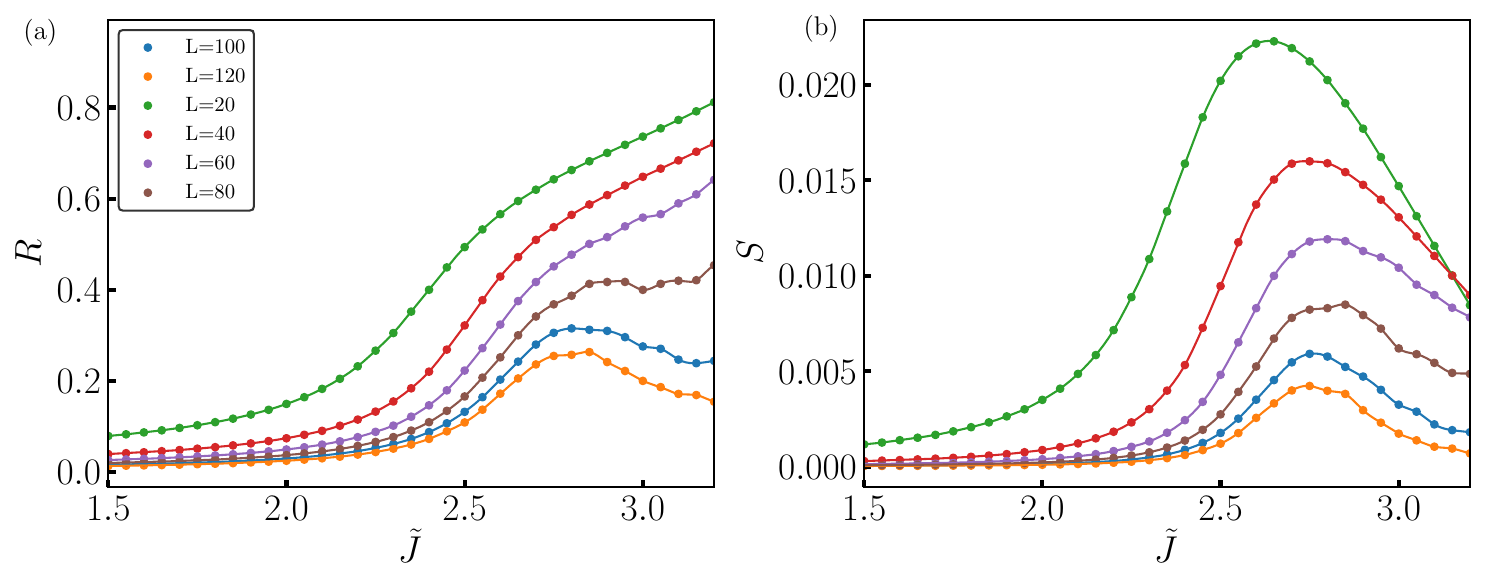}%
\caption{The simulations were performed for the parameter values $\Delta = 0$ and $\tilde K=0.3$. (a) For large $L$ (and nonzero $\tilde{K}$), the static order phase becomes unstable, leading to a droplet-induced swap phase. Spiral defects destroy the droplet-induced swap phase so that $R$ becomes zero. (b) With the increasing system size, the angular-momentum-like order parameter $S$ tends towards zero. This is due to the spiral defects in the system, which destroy the existence of the swap phase. In the thermodynamic limit, there will be complete disorder. \label{6}}
\end{figure}

Our Monte Carlo results at $\Delta = 0$ show that neither the swap phase nor the static ordered phase survives in the two-dimensional NR-BCM in the thermodynamic limit. As illustrated in Fig.~\ref{6}, both the synchronisation order parameter $R$ and the angular-momentum-like order parameter $S$ decrease with system size and approach zero as the lattice length increases. This strongly suggests that, in the thermodynamic limit, the system becomes fully disordered and the apparent ordered phases observed at finite size are not truly stable. Strictly speaking, in 2D, the swap phase occurs through droplet formation (it occurs only in finite systems, since in the thermodynamic limit the swap phase reduces to disorder) and should be called the droplet-induced swap phase \cite{avni2025dynamical}. The actual swap phase, in which the magnetisations are swapped homogeneously, is observed in 3D systems.  In 2D, the swap phase is quickly disrupted by spiral defects, and as a result, the system cannot sustain a persistent, globally coherent swapping of magnetisations. Likewise, the static ordered phase does not truly exist: although local regions may temporarily develop ordered magnetisations, nonreciprocity continually drives species A and B to flip their roles, preventing the system from settling into a stable fixed point. What we observe instead is a droplet-induced swap phase, in which domains repeatedly trade magnetisations with a time period that grows with the interaction strength $\tilde{J}$. However, this behaviour is not indicative of long-range order; in the thermodynamic limit, spiral defects ultimately destroy all kinds of oscillatory order. A similar breakdown of ordered and swap phases in two dimensions has been reported previously in the 2D nonreciprocal Ising model \cite{avni2025nonreciprocal}. However, in higher dimensions, the swap phase is stable, which we also find true for the 3D NR-BCM.

\begin{figure}[h!]
\includegraphics[width=1.0\linewidth]{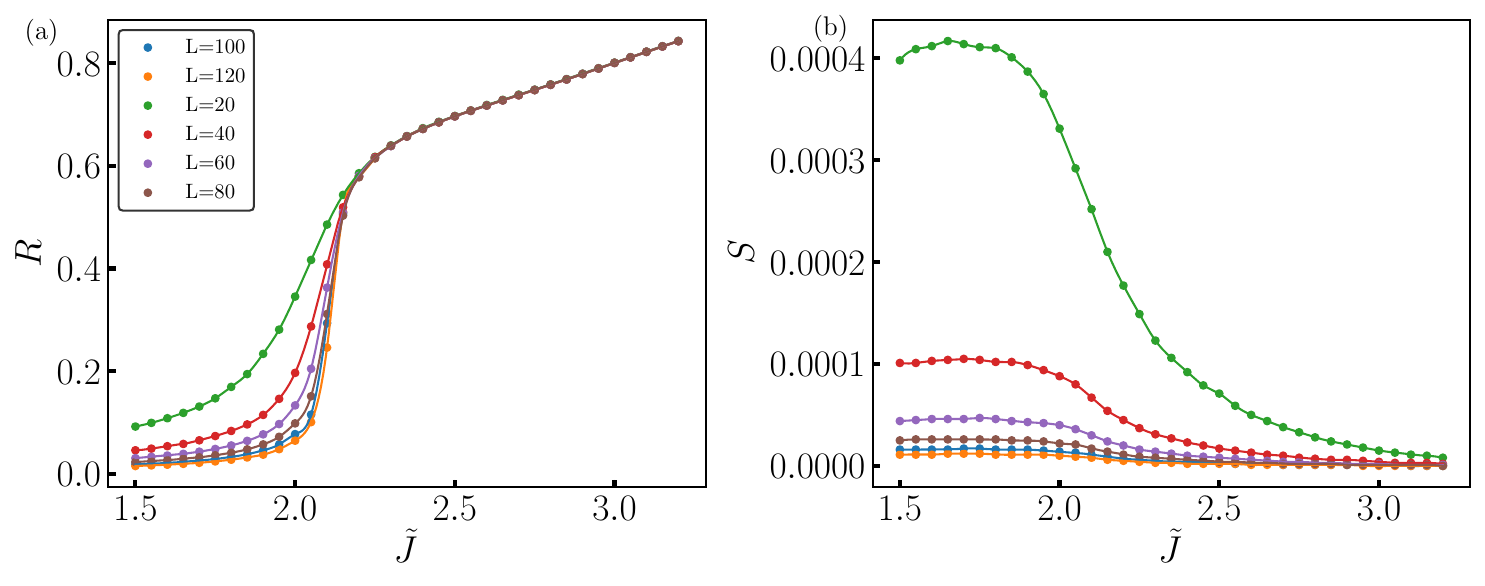}%
\caption{The graphs were plotted for $\Delta = 1.5$, $\tilde K=0.3$. (a) Shows $R$ vs $\tilde J$ from which it is evident that the static order phase is stable. (b) Depicts the graph of $S$ vs $\tilde J$. In the static ordered regime, the angular-momentum-like order parameter $S$ vanishes, confirming that there is no droplet-induced swap phase. \label{7}}
\end{figure}

However, when the Blume–Capel single-ion anisotropy (or chemical potential) \(\Delta\) is nonzero and sufficiently large, the system is able to sustain a stable static order phase. This effect is illustrated in Fig.~\ref{7}, where we plot the order parameters $R$ and $S$ as functions of the interaction strength $\tilde{J}$. Our results show that, as the system size increases, the synchronisation $R$ converges to a nonzero value, indicating the emergence of a robust static order phase in the presence of finite $\Delta$. This demonstrates that introducing a sufficiently strong chemical potential can overcome the disordering effects of nonreciprocity and restore long-range static order in the 2D NR-BCM.

The emergence of static order in the system can be understood by considering the role of the chemical potential $\Delta$.  As $\Delta$ 
increases, it energetically favours species A to occupy the vacant lattice sites more frequently, while the other species is less likely to do so. This imbalance reduces the influence of nonreciprocal interactions on the majority species, as the minority species is no longer able to effectively force the majority spins to flip or reverse their magnetisation. Consequently, the dynamic swapping that characterised the earlier phases is suppressed. With the diminishing effect of nonreciprocity, the system can now settle into a stable static ordered state, allowing long-range magnetisation to persist across the lattice. This mechanism highlights how chemical potential asymmetry can counteract the disordering tendencies induced by nonreciprocal interactions and restore robust order in the 2D NR-BCM.

\begin{figure}[h!]
\includegraphics[width=1.0\linewidth]{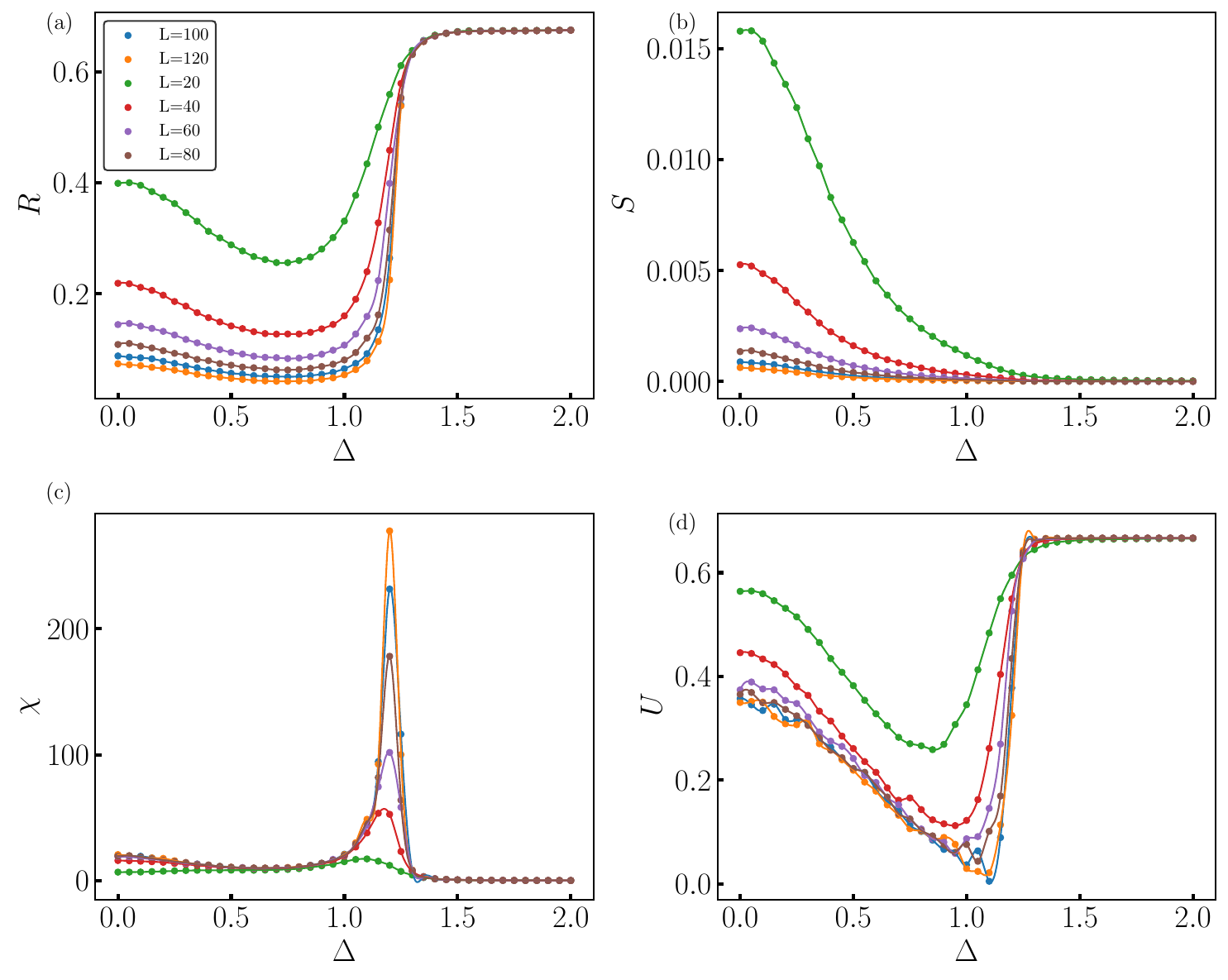}%
\caption{$\tilde J = 2.4$, $\tilde K =0.3$. (a) Before the phase transition, the synchronisation parameter $R$ converges to zero with increasing lattice length $L$. This shows that the transition is from disorder to order in the thermodynamic limit. (b) The swap phase gets destroyed for large values of $L$. The increasing anisotropy parameter $\Delta$ also contributes to the destruction of the swap phase. (c) The maxima of the susceptibility show clear evidence of a phase transition. The order of the phase transition could be discerned from the scaling of its maxima with the lattice length. (d) Here, U is the binder cumulant, and for different lattice lengths, it is seen to intersect at a particular $\Delta$ value. This provides evidence for the transition to be continuous (critical). \label{8}}
\end{figure}

\begin{figure}[h!]
\includegraphics[width=1.0\linewidth]{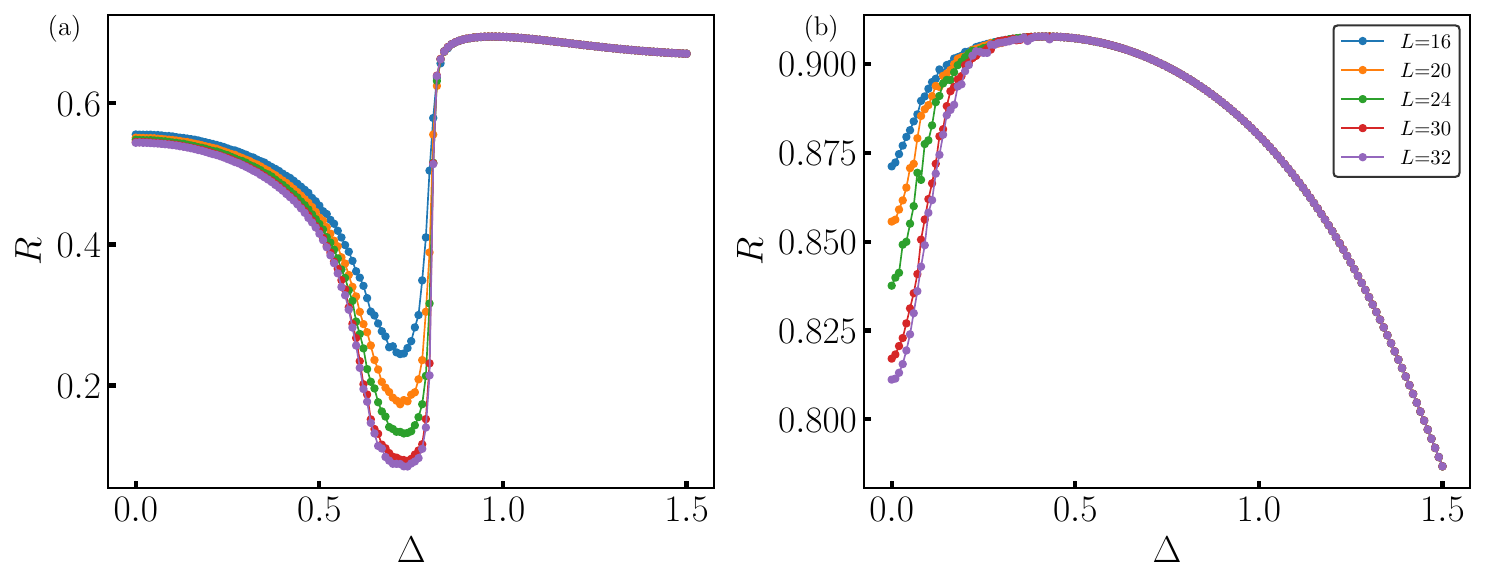}%
\caption{Variation of the synchronisation order parameter $R$ with $\Delta$ for a 3D NR-BCM. The fixed parameter values are taken to be $\tilde J = 2.2$ and $\tilde K = 0.3$. (a) The system does not have a direct transition from swap to static order. Instead, we see, swap $\to$ disorder $\to$ static order. (b) However, with the droplet-induced swap phase, we observe a direct transition to static order. The simulations were performed with $\tilde J = 2.85$ and $\tilde K = 0.3$. \label{9}}
\end{figure}

For lower values of the interaction strength $\tilde{J}$, mean-field theory predicts a direct transition from the swap phase to the static ordered phase as the chemical potential $\Delta$ is varied. However, our Monte Carlo simulations reveal a different scenario: instead of transitioning from the swap phase, the system goes from a disordered state to the static ordered phase, as illustrated in Fig.~\ref{8}(a), (b). In two dimensions, this apparent discrepancy can be partially attributed to the presence of spiral defects, which disrupt long-range order and frustrate the formation of the swap phase. Interestingly, similar behaviour is observed in three-dimensional simulations: even though the swap phase is robust, and spiral defects are absent (see Fig.~\ref{9}(a)), the transition from swap to static is mediated by a region of disorder, indicating that the deviation from mean-field predictions is not solely a consequence of topological defects.
\begin{figure}[h!]
\includegraphics[width=1.0\linewidth]{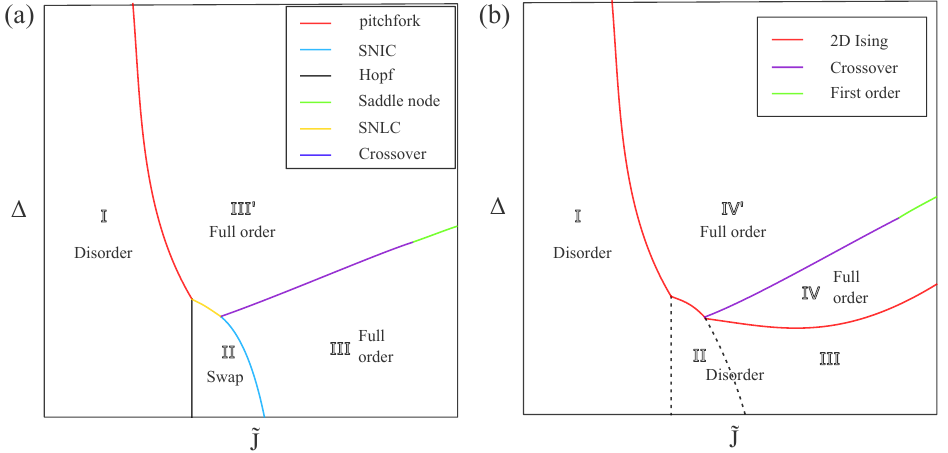}%
\caption{(a) Schematic phase diagram showing the different phases and bifurcations corresponding to the mean-field results. (b) Schematic phase diagram corresponding to the Monte Carlo simulation results at nonzero nonreciprocity. The dashed lines denote the phase transitions which would have existed if the swap and droplet-induced swap phases were not destroyed. \label{10}}
\end{figure}
The ultimate fate of this transition remains unresolved. Thus, the absence of a transition from the swap phase to the static ordered phase in 2D cannot be attributed only to spiral defects. Additional evidence comes from the spin-evolution movie (Supplementary Movie 1 included with this submission as anc/SM1.mp4): in the parameter range close to criticality ($\tilde J = 2.4, \Delta = 1.0, \tilde K = 0.3$)--—where the behaviour appears disordered—--no spiral defects are seen at all. Instead, the system fragments into well-defined magnetisation domains that survive over long times and drift continuously. An analogous behaviour has also been observed in the general asymmetric nonreciprocal Ising model \cite{avni2025nonreciprocal}. However, the transition from the droplet-induced swap phase to the static order in 3D (Fig.~\ref{9}(b)) appears to be direct. The full set of phases realised in the model is summarised in the schematic phase diagram obtained from our Monte Carlo simulations at $\tilde K = 0.3$ (see Fig.~\ref{10}(b)). It is noteworthy that a transition that, in mean-field theory, should manifest as a SNIC---with the oscillation period diverging as the critical point is approached---and the SNLC bifurcation (which is discontinuous) effectively turns into a standard continuous critical transition belonging to the 2D Ising universality class, as we will show later. 

\begin{figure}[h!]
\includegraphics[width=1.0\linewidth]{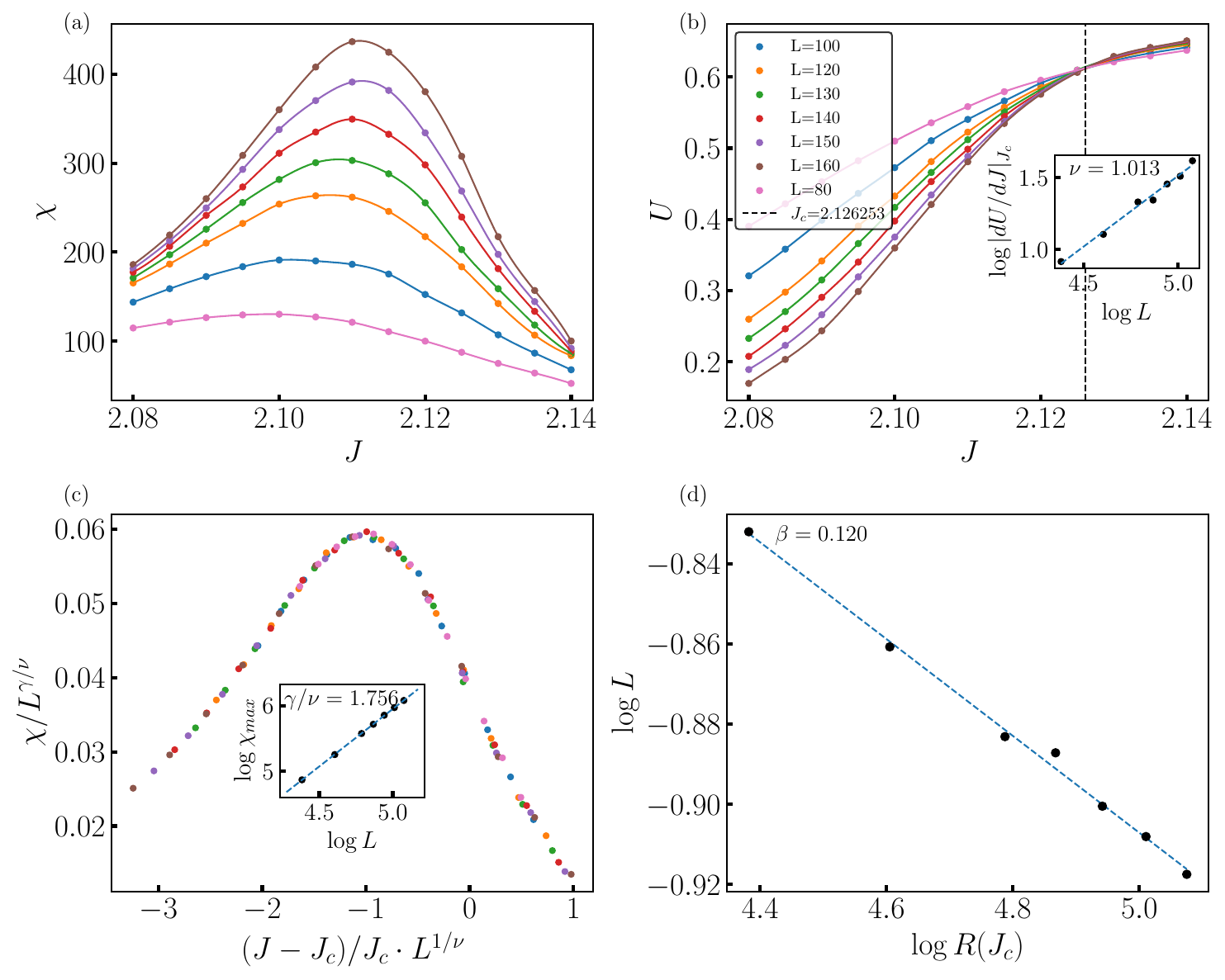}%
\caption{The simulation was performed for the parameter values $\Delta = 1.5$, $\tilde K=0.3$. (a) Susceptibility of the system is plotted for different lattice lengths. (b) Binder cumulant is plotted for different lengths, and the crossing point is identified. The inset shows the scaling of the slope of the binder cumulant at the crossing point with length. From the slope of this scaling, one finds the value of $\nu$. (c) Data collapse at the numerically measured values of the critical exponents $\gamma$ and $\nu$. The inset shows the scaling of the maxima of the susceptibility against length, and its slope yields $\gamma/\nu$. (d) Scaling of the order parameter $R$ (at $\tilde J_c$) with length. The provides the value of the critical exponent $\beta$. \label{11}}
\end{figure}

\begin{figure}[h!]
\includegraphics[width=1.0\linewidth]{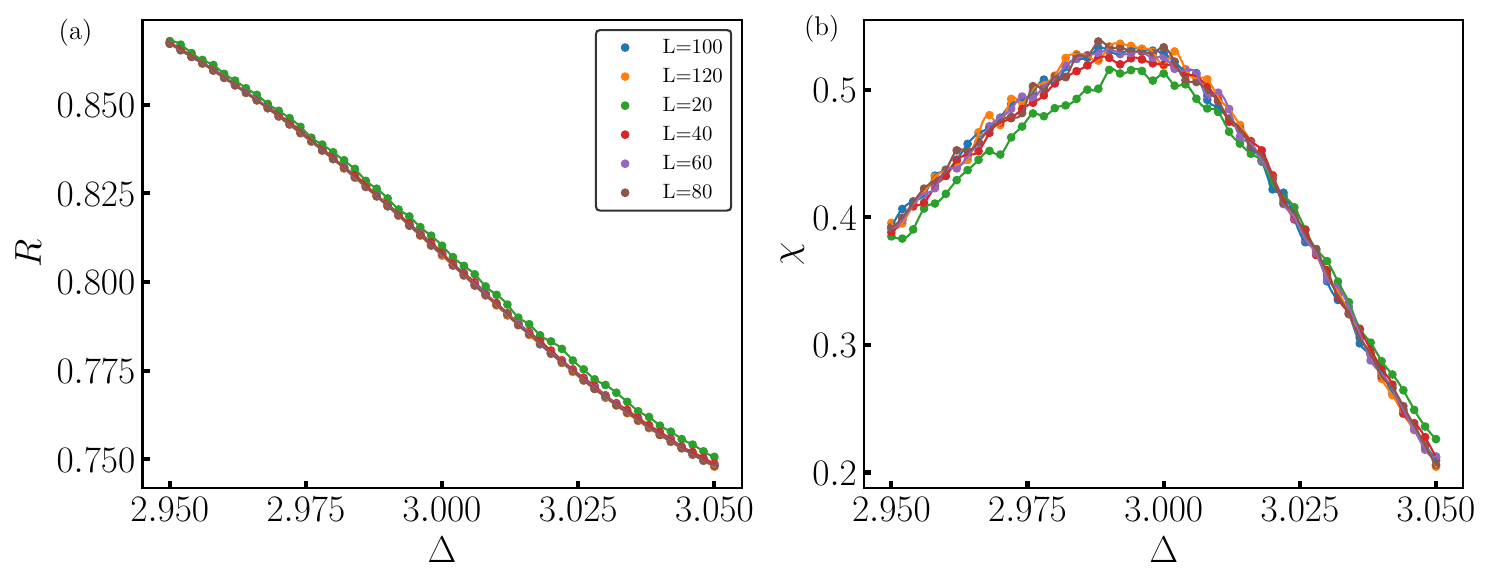}%
\caption{The graphs are for $\tilde J = 5.5$ and $\tilde K = 0.3$. (a) Shows the variation of $R$ with $\Delta$. The transition occurs alongside a continuous change in $R$.
 (b) Susceptibility against $\Delta$ is plotted for different lengths. The peak of the susceptibility is independent of length, indicating a crossover. \label{13}}
\end{figure}

In Fig.~\ref{10}(b), the transition from region II to region IV' corresponds to a continuous (critical) phase transition from a disordered state—where the swap phase is destroyed by spiral defects—to a statically ordered state. This transition falls within the 2D Ising universality class. Moreover, mean-field theory predicts that region III should exhibit full order, but Monte Carlo simulations instead reveal a droplet-induced swap phase that is disrupted by spiral defects, preventing the establishment of long-range order and resulting in disorder. However, as discussed previously, for sufficiently large values of $\Delta$, the system is able to support a fully ordered phase, denoted as region IV. Consequently, the transition from region III to region IV can be understood as a shift from disorder to a fully ordered state, and this transition is also found to be second-order and consistent with the 2D Ising universality class. The transition from region IV to region IV' (both regions represent the same phases), on the other hand, exhibits more nuanced behaviour. For lower values of $\tilde{J}$, the change appears to be a smooth crossover, as evidenced by the negligible peak in the susceptibility (see Fig.~\ref{13}) and the nearly identical results across different system sizes. In contrast, for higher values of $\tilde{J}$, this transition becomes first-order. This is similar to the results we got from the mean-field analysis. Here, we can say that the region~IV and IV' are analogous to the gas and liquid phases in terms of phase transition. 

\begin{figure}[h!]
\includegraphics[width=1.0\linewidth]{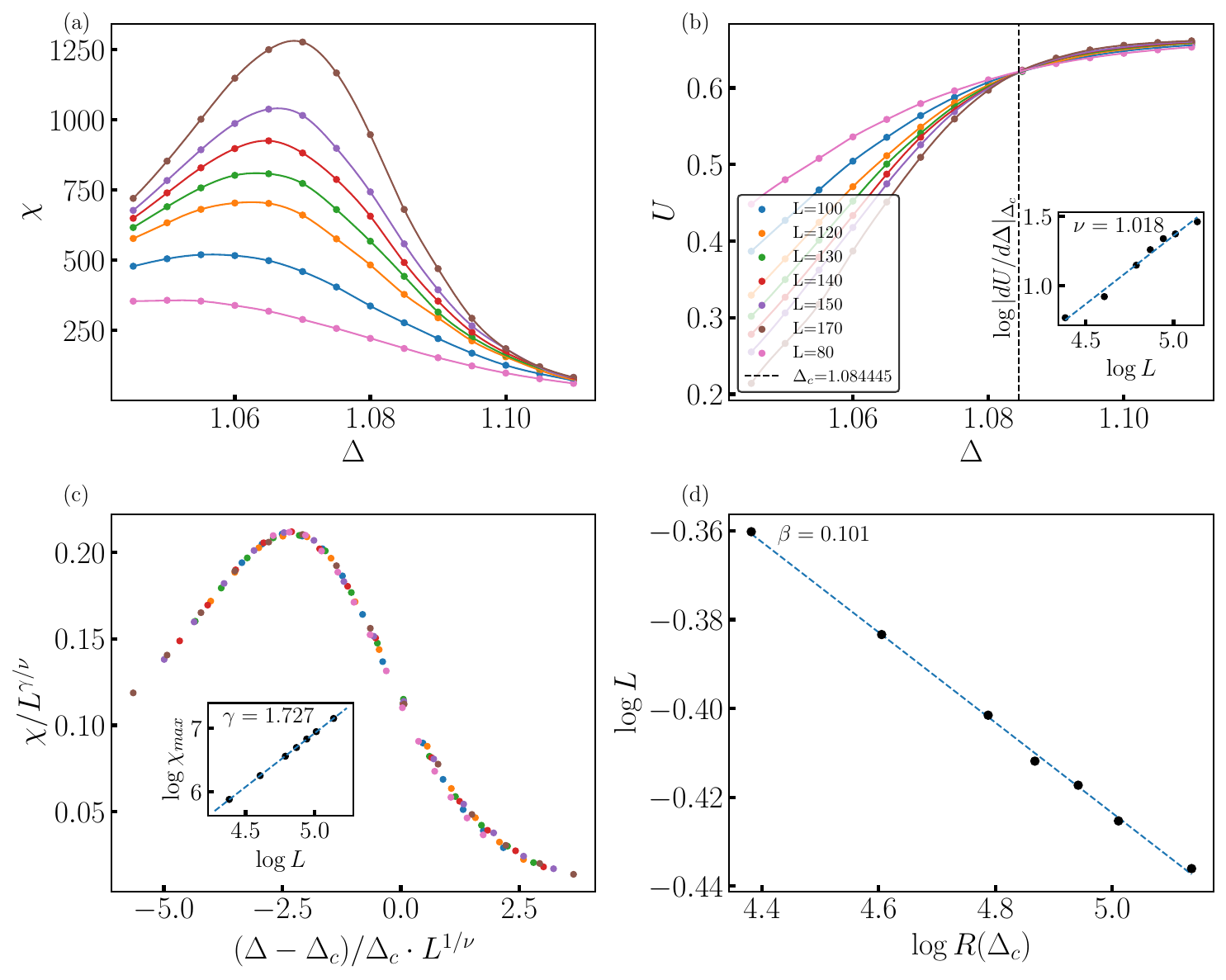}%
\caption{The results are plotted for $\tilde J = 2.9$, $\tilde K=0.3$. (a) Illustrates susceptibility against $\Delta$ for different values of length. (b) Binder cumulant crossing for different lengths. The inset shows the scaling of the slope of the binder cumulant at the crossing vs length. (c) Data collapse at the numerically measured values of the critical exponents $\gamma$ and $\nu$. The inset shows the scaling of the maxima of the susceptibility with length. (d) Scaling of the order parameter $R$ (at $\Delta_c$) with length. The provides the value of the critical exponent $\beta$. \label{12}}
\end{figure}

To determine whether a given phase transition is continuous (critical) and to extract the corresponding critical exponents, we perform a finite-size scaling (FSS) analysis. According to FSS theory, the peak susceptibility scales as $\chi_{max}\sim L^{\gamma/\nu}$ for second-order transitions, where $\chi = L^d(\langle R^2\rangle - \langle R\rangle^2)$ \cite{fisher1972scaling}. In contrast, for first-order transitions, the scaling follows $\chi_{max}\sim L^d$ \cite{challa1986finite}. Thus, by examining the slope of a log–log plot of $\chi_{max}$ vs $L$, we can identify the order of the transition. Additional evidence for the order of the transition is provided by the Binder cumulant, defined as $U_L = 1 - \langle R^4\rangle/3\langle R^2\rangle^2$. In a first-order transition, $U_L$exhibits a sharp dip at the transition point, reflecting the coexistence of competing phases. By contrast, for a continuous (second-order) transition, $U_L$ evolves smoothly across the critical region, and curves corresponding to different system sizes intersect at a common point, $\tilde J_c(\infty)$, where $\tilde J_c(\infty)$ is the critical interaction strength in the thermodynamic limit.

We begin by analysing the transition from region II to region IV'. To this end, we fix the model parameters at $\Delta = 1.5$, $\tilde{K} = 0.3$ and vary the coupling strength $\tilde J$. The finite-size scaling of the susceptibility peak reveals a slope of $\gamma/\nu = 1.756 \pm 0.007$ (see inset Fig.~\ref{11}(c)). Since this value is strictly less than the spatial dimension $d$ = 2, the transition is clearly continuous (critical). Using standard scaling relations, we further extract the critical exponents, obtaining $\gamma = 1.779 \pm 0.084$, $\nu=1.013 \pm 0.047$, and $\beta = 0.120 \pm 0.003$ (Fig.~\ref{11}(b), (d)). A successful data collapse using these exponents, shown in Fig.~\ref{11}(c), provides additional confirmation of their accuracy. Collectively, these results demonstrate that the transition from disorder to static order (II $\to$ IV') in the 2D NR-BCM is a second-order phase transition belonging to the Universality class of the two-dimensional Ising model.

The transition from region III to region IV corresponds to a switch from a disordered state—where the droplet-induced swap phase is disrupted by spiral defects—to a fully ordered phase. To characterise this transition, we perform a finite-size scaling analysis, with the results shown in Fig.~\ref{12}. By examining the plot of $\log{\chi_{max}}$ vs $\log{L}$, we find a slope of $\gamma/\nu = 1.695 \pm 0.006$, indicating that the transition is second-order. From the scaling analysis, the critical exponents are determined to be $\nu = 1.018 \pm 0.057$, \textbf{$\gamma = 1.727 \pm 0.097$} and  $\beta = 0.101 \pm 0.001$. The values of the exponents are very close to the 2D Ising model. Hence, they belong to the 2D Ising universality class. Furthermore, the data collapse illustrated in Fig.~\ref{12}(c) provides validation of these critical exponents.

\begin{figure}[h!]
\includegraphics[width=1.0\linewidth]{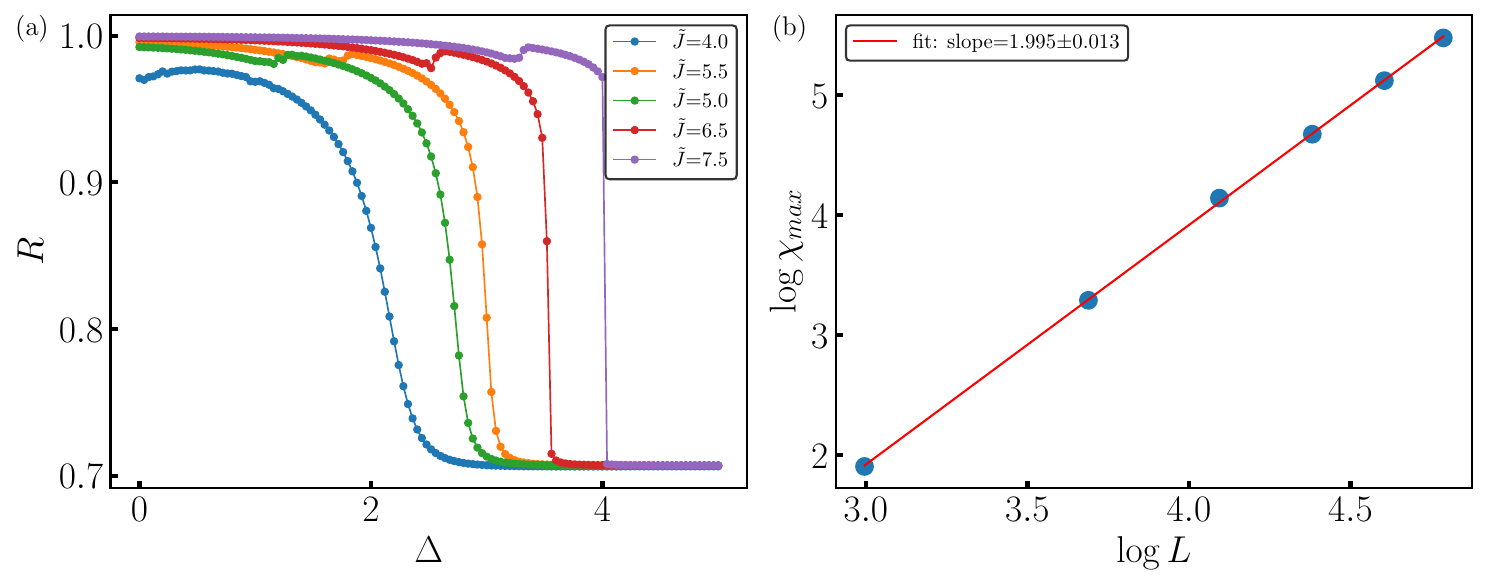}%
\caption{(a) Plot of $R$ with $\Delta$ for different interaction strengths $\tilde J$. The nonreciprocity parameter is taken to be $\tilde K = 0.3$. For increasing $\tilde J$, the transition becomes sharper, which is an indicator of first-order transitions at large values of $\tilde J$. (b) Scaling of the peak of the susceptibility with length at $\tilde J = 7.5$. The slope is $\approx 2$, which is a definite indication of a first-order phase transition. Finite size scaling was performed with lattice lengths $L = 20, 40, 60, 80, 100, 120$\label{14}}
\end{figure}

The 2D Monte Carlo simulation results indicate that the transition from region IV to region IV' represents a crossover. As shown in Fig.~\ref{13}, the order parameter $R$ varies smoothly without any discontinuity, and the corresponding susceptibility remains essentially independent of system size. However, as $\tilde J$ increases, the transition from IV to IV' becomes progressively sharper (see Fig.~\ref{14}(a)), indicating the onset of first-order behaviour at large $\tilde J$.
In the phase diagram, this is represented by the green line (Fig.~\ref{10}(b)). Mean-field theory similarly predicts the occurrence of first-order phase transitions at high $\tilde J$ which are reflected as discontinuities in the colour map for $\tilde{J} > 4.0$ (see Fig.~\ref{1}(c)). Our Monte Carlo simulations further reveal that for approximately $\tilde J > 7.0$ the system begins to exhibit clear signatures of a first-order transition, qualitatively consistent with the mean-field predictions, including abrupt changes in the order parameter. This is supported by the FSS analysis at $\tilde J = 7.5$ and $\tilde K = 0.3$ shown in Fig.~\ref{14}(b), which shows that the susceptibility maxima scale with the length as $\sim L^{1.995\pm0.013}$. This highlights how the interaction strength strongly influences the nature of the transitions in the 2D NR-BCM. Moreover, the line of first-order phase transitions terminates at a critical point, analogous to the one in the liquid–gas–solid phase diagram.  

\begin{figure}[h!]
\includegraphics[width=1.0\linewidth]{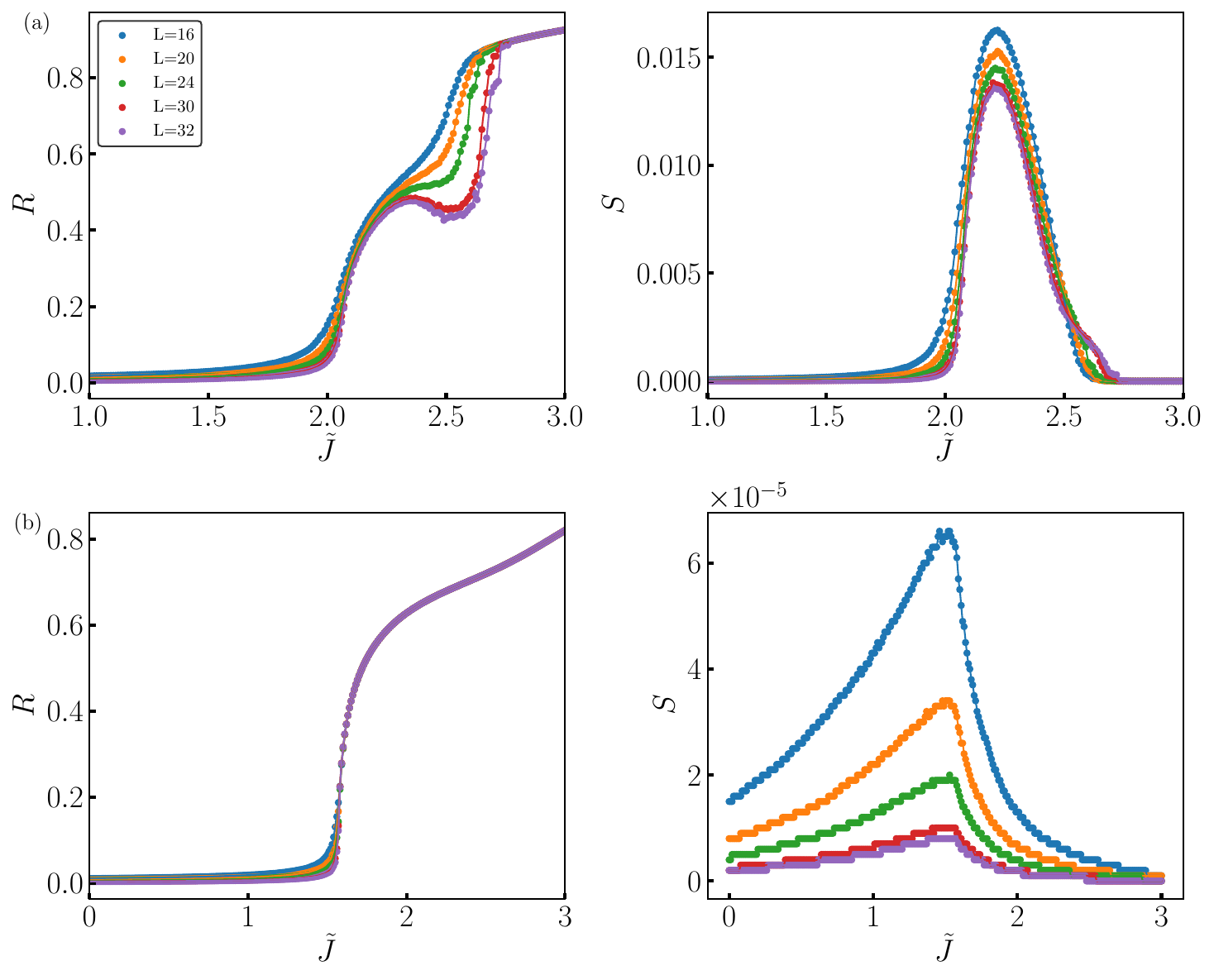}%
\caption{(a) Shows the graphs of order parameter $R$ and $S$ with respect to $\tilde J$ at $\Delta = 0.5$ and $\tilde K = 0.3$. The swap phase is stable in 3D as it is seen to converge at a nonzero $S$ with increasing lattice length. The sequence of phase transitions with increasing $\tilde J$ is disorder $\to$ swap $\to$ droplet-induced swap. (b) At $\Delta = 1.5$, the system undergoes a phase transition from disorder $\to$ full order. Here, we don't observe the swap phase, which is qualitatively similar behaviour as the mean-field result. \label{15}}
\end{figure}

Fig.~\ref{15} illustrates the presence of multiple phases in the 3D NR-BCM, which includes a disordered state, stable swap phase, and static order phase. As shown in Fig.~\ref{15}(a), for $\Delta$ = 0.5, the system exhibits a stable swap phase, indicated by the convergence of the order parameter $S$ to a nonzero value as the lattice size increases. In this regime, no static order is observed; instead, the dynamics are dominated by a droplet-induced swap phase. At a higher chemical potential, $\Delta$ = 1.5 (Fig.~\ref{15}(b)), the swap phase disappears, and the system undergoes a transition from a disordered state to an ordered state as $\tilde J$ is increased, followed by a smooth crossover. Overall, the 3D Monte Carlo simulations qualitatively reproduce the predictions of mean-field theory. However, a notable difference is the transition from the swap phase to the static order phase. As observed, the transition occurs indirectly—unlike in the mean-field analysis, where a SNIC or SNLC bifurcation is predicted. A further key difference lies in the behaviour of the droplet-induced swap phase: it remains uncertain whether this phase will eventually evolve into a disordered state in the thermodynamic limit. Using mean-field analysis, we can only show the existence of the droplet-induced swap phase. 

\section{Conclusion}

We have introduced and studied a minimal two-species NR-BCM, combining a systematic mean-field bifurcation analysis with extensive Monte Carlo simulations in both two and three spatial dimensions. The mean-field approach categorises the system into a small set of dynamical regimes—disorder, time-dependent swap, and static order—separated by Hopf, SNIC, SNLC, pitchfork, and saddle-node bifurcations. However, our numerical simulations show that in low-dimensional systems, this mean-field framework is significantly modified by the presence of fluctuations and topological defects, which reshape the phase behaviour and the nature of transitions, revealing richer and more complex dynamics than predicted by mean-field theory alone.

A key finding of our study is that species-dependent vacancy energetics, which enters into the model as a chemical-potential imbalance ($\Delta_A=-\Delta_B$), serves as a powerful control parameter in the system. A finite single-ion anisotropy $\Delta$ effectively suppresses the nonreciprocal swap dynamics and promotes the stabilisation of static order. In two dimensions, spiral and droplet defects generally disrupt long-lived swap behaviour unless $\Delta$ is sufficiently large to bias the occupation of vacancies; when static order is restored, the resulting disorder-to-order transitions exhibit critical behaviour consistent with the 2D Ising universality class. Moreover, within the static ordered phase, we observe a crossover that sharpens into a line of first-order phase transitions; these two regimes are separated by a critical point, analogous to the termination of the liquid–gas coexistence curve. 

In contrast, three-dimensional simulations retain many features predicted by mean-field theory, including a robust swap phase at small $\Delta$. Nonetheless, the pathway from swap to static order is indirect—proceeding via swap $\to$ disorder $\to$ static—and the ultimate fate of droplet-mediated swap regions in the thermodynamic limit remains an open question, highlighting the subtle interplay between nonreciprocity, dimensionality, and vacancy energetics.

Several promising directions emerge from this work. First, larger-scale numerical simulations and finite-size scaling analyses adapted to the kinetics of droplet formation are needed to clarify the thermodynamic-limit behaviour of the droplet-induced swap regime. Determining the precise location of the critical point (the point where the line of first-order phase transition ends) and identifying its associated universality class also remain open challenges. Additionally, extending the model to incorporate unequal chemical potentials, conserved dynamics, longer-range interactions, or active stochastic forcing could provide deeper insight into the system’s behaviour. As discussed previously, asymmetric chemical potentials give rise to novel behaviours, such as a swap phase occurring around a nonzero $(M_A, M_B)$. A more detailed investigation of this phenomenon is required, using both mean-field analysis and Monte Carlo simulations. Finally, exploring connections to experimental platforms—such as active colloids, driven magnetic materials, parametrically driven oscillator arrays, or robotic metamaterials—would allow testing the robustness and practical applicability of the control mechanism uncovered in this study.

In summary, the NR-BCM introduced in this work offers a minimal yet versatile framework that illustrates how local vacancy energetics can suppress nonreciprocal, time-dependent dynamics while simultaneously restoring static universal order. At the same time, it highlights the crucial influence of defects in determining the non-equilibrium phase behaviour of the system. These results pave the way for a deeper theoretical understanding of nonreciprocal many-body systems and provide a foundation for the practical design and control of such systems in experimental settings.

\section{Acknowledgments}
The authors acknowledge financial support from the
Department of Atomic Energy, India through the plan
Project (RIN4001-SPS).
\bibliography{references}

\end{document}